\documentclass[runningheads]{llncs}
\usepackage[T1]{fontenc}
\usepackage{mwe} 
\usepackage{multirow}
\usepackage{siunitx}
\usepackage{tikz}
\usepackage{pgfplots}

\usepackage{arydshln}
\usepackage{amsmath}
\usepackage{amssymb}
\usepackage[colorlinks, citecolor=green]{hyperref}
\usepackage{adjustbox}
\usepackage{subcaption}
\usepackage{booktabs}
\usepackage{hhline}
\usepackage{svg}
\usepackage{graphicx,verbatim}
\usepackage{caption}
\usepackage{tikz} 
\newcommand{\supersmall}{\fontsize{7}{8.4}\selectfont}
\definecolor{uvitbg}{HTML}{E3F2FD} 
\definecolor{ldmbg}{HTML}{FCE4EC}  
\definecolor{ditbg}{HTML}{E8F5E9}  
\definecolor{mdtbg}{HTML}{FFF3E0}  

\begin{document}
\title{Reconstruct or Generate: Exploring the Spectrum of Generative Modeling for Cardiac MRI}
%

\author{
Niklas Bubeck\inst{1,2} \and
Yundi Zhang\inst{1,3} \and
Suprosanna Shit\inst{4} \and 
Daniel~Rueckert\inst{1,2,3,5} \and
Jiazhen Pan\inst{1,3}
}

\authorrunning{N. Bubeck et al.}
%
\institute{School of Computation, Information and Technology, Technical University of Munich, Germany
\and Munich Center for Machine Learning, Technical University of Munich, Germany
\and School of Medicine, Klinikum Rechts der Isar, Technical University of Munich, Germany
\and Department of Quantitative Biomedicine, University of Zurich, Switzerland
\and Department of Computing, Imperial College London, UK
\\
\email{\{niklas.bubeck,jiazhen.pan\}@tum.de}
}

\titlerunning{Exploring the Spectrum of Generative Modeling for Cardiac MRI}
    
\maketitle 

\begin{abstract}
In medical imaging, generative models are increasingly relied upon for two distinct but equally critical tasks: reconstruction, where the goal is to restore medical imaging (usually inverse problems like inpainting or superresolution), and generation, where synthetic data is created to augment datasets or carry out counterfactual analysis. Despite shared architecture and learning frameworks, they prioritize different goals: generation seeks high perceptual quality and diversity, while reconstruction focuses on data fidelity and faithfulness. In this work, we introduce a "generative model zoo" and systematically analyze how modern latent diffusion models and autoregressive models navigate the reconstruction-generation spectrum. We benchmark a suite of generative models across representative cardiac medical imaging tasks, focusing on image inpainting with varying masking ratios and sampling strategies, as well as unconditional image generation. Our findings show that diffusion models offer superior perceptual quality for unconditional generation but tend to hallucinate as masking ratios increase, whereas autoregressive models maintain stable perceptual performance across masking levels, albeit with generally lower fidelity.

\end{abstract}


\section{Introduction}

Generative models have gained increasing prominence in medical imaging, offering powerful solutions for a range of tasks such as image reconstruction~\cite{xie2022measurement,chung2022score,peng2022towards}, inpainting~\cite{rouzrokh2022multitask}, enhancement~\cite{gong2024pet,hu2022unsupervised}, and data augmentation~\cite{pinaya2022brain}. Compared to natural image applications, however, clinical scenarios demand far greater emphasis on preserving anatomically and diagnostically essential details. In particular, methods must balance two competing goals: fidelity—capturing subtle, subject-specific features measured by metrics such as PSNR or SSIM—and perception—preserving the “naturalness” of rendered images often evaluated through metrics like Frechet Inception Distance (FID). Achieving both goals is challenging given the stakes of clinical decision-making, where distorted or “hallucinated” structures can lead to harmful outcomes.

\noindent Most contemporary approaches to medical generative modeling adopt a two-stage architecture: a first-stage encoder–decoder projects images into a latent space and back, while a second-stage prior operates on these latent codes to reconstruct or generate images. Although these solutions work well in narrowly defined tasks with either strong conditioning information (e.g., near-complete scans for reconstruction) or no input at all (e.g., unconditional synthesis), clinical tasks often occupy a middle ground. Partial or imperfect inputs—such as images with small or large regions missing—require the model to rely selectively on learned priors. Yet, how to optimally navigate this reconstruction–generation continuum remains an open question.

\noindent Prior works like Blau et al.~\cite{blau2018perception} have discussed the distortion-perception tradeoff for image restoration, showing that improving perceptual quality inevitably leads to higher distortion, thus decreasing fidelity. Yao et al.~\cite{yao2025reconstruction} discuss the reconstruction-generation tradeoff solely within the first stage model, showing that increasing tokenizer dimension improves detail but harms generation, and address this with a VA-VAE that aligns latents with vision foundation models to expand the tradeoff frontier. In this work, we argue that the dichotomy between reconstruction of undersampled data and unconditional generation is better understood as a \textbf{continuous spectrum} in the medical domain. Therefore, our evaluation is focused on the second stage, by systematically varying the amount of available conditioning information by masking different portions of the input image. Through this analysis, we provide new insights into the interplay between data-driven priors and training paradigms in high-stakes medical settings.

\noindent We introduce a modular “generative models zoo” to address diverse medical imaging needs. In the first stage, we include commonly used compression encoders and decoders—such as variational autoencoders (VAEs)~\cite{kingma2013auto} or vector-quantized GANs (VQ-GANs)~\cite{esser2021taming}—that reduce spatial redundancy and capture relevant anatomical features. In the second stage, we curate a variety of generative priors, including diffusion-based~\cite{rombach2022high,peebles2023scalable,bao2023all,dhariwal2021diffusion,gao2023masked} and autoregressive models~\cite{chang2022maskgit,esser2021taming,li2023mage}. Through this flexible framework, users can readily mix and match different first- and second-stage components to tailor performance for specific clinical goals.

\noindent We further compare diffusion-based and autoregressive models on 2D cardiac MRI, assessing reconstruction and generation across perception, fidelity, and complexity metrics. Diffusion models excel at fidelity with sufficient context but hallucinate under moderate masking, while autoregressive models offer more stable perceptual quality and better alignment at high masking ratios. Among them, MaskGIT~\cite{chang2022maskgit} achieves a strong balance—delivering high-fidelity reconstructions at low masking and diverse, realistic outputs when context is sparse—highlighting its potential for clinical applications.

\noindent In summary, our contributions are threefold:
\begin{enumerate}
    \item We recast medical reconstruction and generation as points along a \emph{continuous spectrum rather than a binary distinction}, revealing that perception can be maximized at both extremes of the masking ratio.
    \item We introduce a \emph{“generative models zoo”} for medical image generation, modularizing first-stage compression (VAEs, VQ-GANs) and second-stage models (diffusion-based and autoregressive models) into an easily extensible framework.
    \item We conduct a \emph{comprehensive evaluation of various architectures}—comparing fidelity, perception, and computational efficiency under different masking ratios—and demonstrate that MaskGIT-like architectures emerge as the most robust and versatile approach across the reconstruction–generation spectrum.
\end{enumerate}
 
\section{Methods}

\begin{figure}[t]
    \centering
\includegraphics[width=\textwidth]{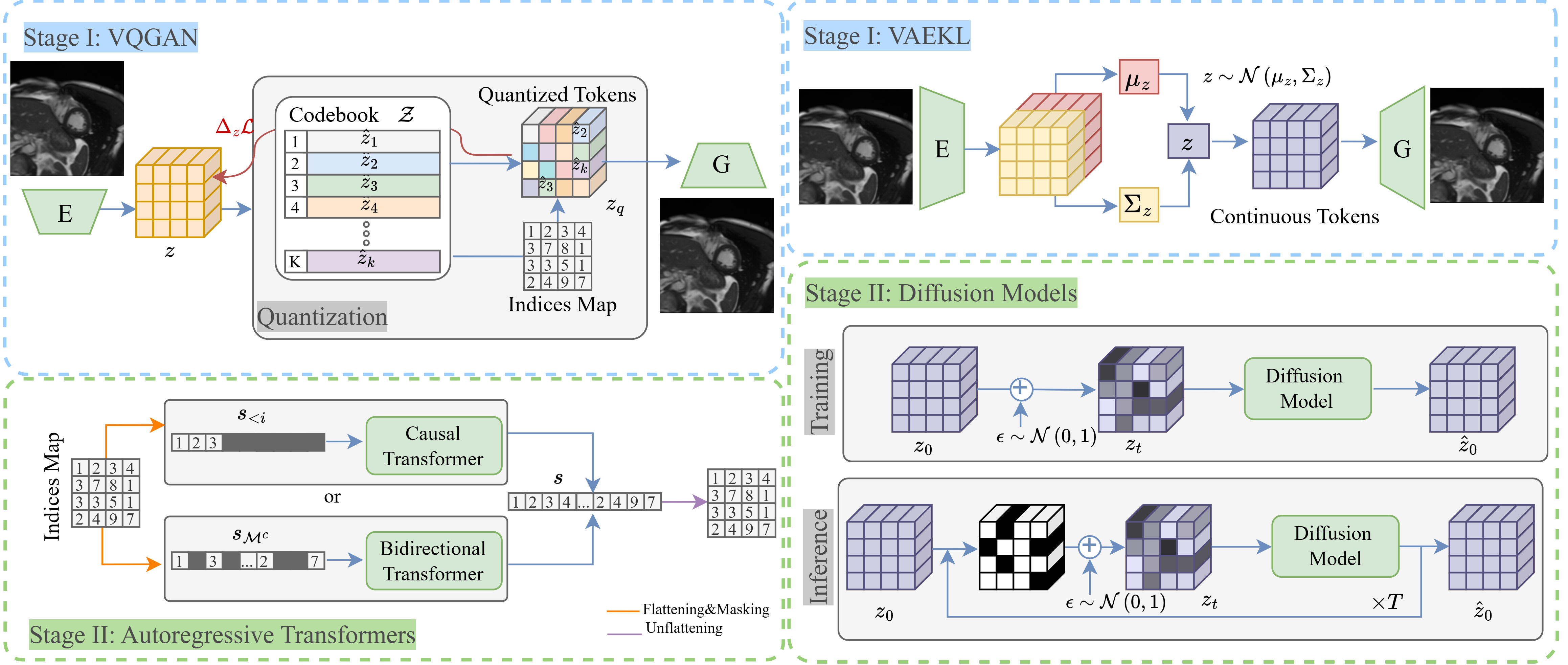}
    \caption{\textit{\textbf{Overview of our two stage approach:}} For the VQGAN (top left), images are encoded into discrete latent tokens via vector quantization. These tokens are then modeled autoregressively (bottom left) using either causal or bidirectional Transformers. Alternatively, the VAE (top right) encodes images into continuous latent variables. A diffusion model (bottom right) is employed in the second stage to iteratively denoise the latent representation, optionally performing known-region injection for inference.}

    \label{fig:main_figure}
\end{figure}

We employ a two-stage framework as shown in Fig.~\ref{fig:main_figure} for image generation and reconstruction. In the first stage, a pretrained autoencoder compresses images into a latent space, which is either continuous (as in VAEs) or discrete (as in VQ-GANs). The second stage uses generative models—either diffusion-based or autoregressive models—to synthesize or reconstruct images from these latent representations.

\subsection{Stage I: Tokenization}
\noindent\textit{\textbf{Variational Autoencoder.}}
Utilizing a VAE trained from scratch, we map an input image \(x\) to a continuous latent variable $z$ via an encoder $E$ approximating the posterior $q_\phi(z|x)$ with $z \sim \mathcal{N}\left(\mu_z, \Sigma_z\right)$. From this latent representation the image can be reconstructed via a decoder $G$ estimating the likelihood \(p_\theta(x|z)\). The model is then trained by maximizing the Evidence Lower Bound (ELBO):

\begin{equation}
\mathcal{L}(\theta, \phi; x) = -\mathrm{D_{KL}}\big(q_\phi(z|x)\,||\,p_\theta(z)\big) + \mathbb{E}_{q_\phi(z|x)}\big[\log p_\theta(x|z)\big]
\end{equation}

\noindent where \(\mathrm{D_{KL}}\) is the Kullback-Leibler divergence and $p(z)=\mathcal{N}(0, I)$ is the prior defined as a Gaussian by design.

\noindent\textit{\textbf{VQGAN.}} 
Alongside the VAE, we train a VQGAN~\cite{van2017neural}, which replaces the continuous latent space with a discrete codebook \( \mathcal{Z} = \left\{ \hat{z}_k \right\}_{k=1}^K \) of size $K$. The encoder \( E \) maps an input \( x \) to a continuous latent representation \( z \), which is then quantized to $z_q$ by minimizing the difference to the closest entry in the codebook:
$z_q = \underset{\hat{z}_k \in \mathcal{Z}}{\arg\min} \left\| z - \hat{z}_k \right\|^2_2.$
\noindent The decoder \( G \) reconstructs the input from the quantized latent \( z_q \), modeling the likelihood \( p_\theta(x \mid z_q) \). Overall, the training objective consists of a reconstruction loss, a codebook loss, and a commitment loss:

\begin{equation}
\mathcal{L}_{\text{VQ}} = \| x - \hat{x} \|_2^2 + \| \mathrm{sg}[z] - z_q \|_2^2 + \beta \| z - \mathrm{sg}[z_q] \|_2^2,
\end{equation}

\noindent where \( \mathrm{sg}[\cdot] \) denotes the stop-gradient operator, used to apply the straight-through estimator by allowing gradients \( \Delta_z \mathcal{L} \) to backpropagate as if quantization were the identity function.
\( \beta \) controls the strength of the commitment loss. To enhance the perceptual quality of the reconstructions, VQGAN incorporates a generative adversarial network (GAN) into the training process. A discriminator $D$ is jointly trained to distinguish between real images  $x$ and reconstructions $\hat{x} = G(z_q)$. Following the patch-based approach introduced by Isola et al.~\cite{isola2017image}, the discriminator operates on local patches rather than the entire image. The adversarial losses for the generator and discriminator are defined as:
\begin{align}
\mathcal{L}_{\text{GAN}}^{\text{G}} &= -\mathbb{E}_{x} \left[ \log D(\hat{x}) \right], &
\mathcal{L}_{\text{GAN}}^{\text{D}} &= -\mathbb{E}_{x} \left[ \log D(x) \right] - \mathbb{E}_{x} \left[ \log (1 - D(\hat{x})) \right].
\end{align}
\noindent The final objective for training VQGAN combines the VQ loss with the adversarial component:
\begin{equation}
\mathcal{L}_{\text{VQGAN}} = \mathcal{L}_{\text{VQ}} + \lambda \mathcal{L}_{\text{GAN}}^{\text{G}},
\end{equation}

\noindent where \( \lambda \) is a weighting factor balancing diversity against fidelity.

\subsection{Stage II: Generation}
\noindent\textit{\textbf{Diffusion Models.}} 
As generative models, we employ and compare several variants of latent diffusion models within the Denoising Diffusion Probabilistic Models (DDPM) framework~\cite{ho2020denoising,song2020score}. DDPMs define a generative process by learning to invert a fixed Markovian forward process that gradually adds Gaussian noise to the latents of the VAE. This forward process transforms a clean latents $z_0$ into a sequence of progressively noisier versions $z_t$, for $ t = 1, \dots, T $, where each $z_t$ is sampled via the following marginal distribution:
\begin{equation}
    q(z_t \mid z_0) = \mathcal{N}(\sqrt{\bar\alpha_t}z_0, (1 - \bar\alpha_t)\mathbf{I}),
\end{equation}
with $\bar\alpha_t $ denoting the preserved fraction of the original signal at timestep $t$.

\noindent To recover the data distribution, a neural network \( \epsilon_\theta(z_t, t) \) is trained to predict the noise component $ \epsilon $ that was added to the data at each timestep. The model is optimized using the objective:
\begin{equation}
    \mathcal{L}_{\text{DDPM}} = \mathbb{E}_{x_0, \epsilon, t} \left[ \| \epsilon_\theta(x_t, t) - \epsilon \|^2_2 \right].
\end{equation}
Sampling from the model is then performed by reversing the noising process in a step-wise manner, using the following formulation:
\begin{align}\label{equation:reverse_diffusion_temp}
    z_{t-1} =\ & \sqrt{\bar\alpha_{t-1}} \left( \frac{x_t - \sqrt{1 - \bar\alpha_t} \epsilon_\theta(z_t, t)}{\sqrt{\bar\alpha_t}} \right) \nonumber \\
    & + \sqrt{1 - \bar\alpha_{t-1} - \sigma_t^2} \cdot \epsilon_\theta(z_t, t) + \tau \cdot \sigma_t \epsilon_t,
\end{align}

\noindent where \( \epsilon_t \sim \mathcal{N}(0, \mathbf{I}) \) is noise and \( \sigma_t \) is the timestep-dependent variance. A temperature parameter \( \tau \) modulates sample diversity: \( \tau = 1 \) recovers standard DDPM sampling, \( \tau < 1 \) sharpens outputs, and \( \tau > 1 \) increases diversity. Final reconstructions \( \hat{x} \) are decoded via the VAE.

\noindent For inpainting, we make use of known-region injection for all models, denoising only the masked regions while re-injecting known latents at each diffusion step to ensure consistency with the input. 

\noindent\textit{\textbf{Autoregressive Models.}}  
We use autoregressive models to learn the distribution over discrete latent indices produced by a VQGAN encoder \( E \). Given an image \( x \), the encoder maps it to a sequence \( s = (s_1, \ldots, s_n) \) of length $n$, where each index \( s_i \in \{1, \ldots, K\} \) corresponds to a codebook vector \( \hat{z}_{s_i} \in \mathcal{Z} \). This sequence forms a compressed representation that serves as the Transformer’s modeling target. In the standard autoregressive (AR) setup (e.g., Taming Transformers~\cite{esser2021taming}), the Transformer models the joint token distribution $p(s)$ via a causal factorization, and is trained to minimize it's negative log-likelihood $\mathcal{L}_{\text{AR}}$:
\begin{align}
  p(s) = \prod_{i=1}^{n} p(s_i \mid s_{<i}), \quad
  \mathcal{L}_{\text{AR}} = \mathbb{E}_{x \sim p(x)} \left[ -\log p(s) \right].
\end{align}

\noindent During sampling, we refine \( p(s_i \mid s_{<i}) \) using temperature scaling and top-\(k\) filtering. Temperature \( \tau \) scales logits as \( \ell / \tau \), where lower \( \tau \) sharpens the distribution and higher \( \tau \) increases randomness. Top-\(k\) limits sampling to the \( k \) most probable tokens clamping the tails of the distribution.

\noindent In masked-autoregressive (MAR) variants (e.g., MaskGIT~\cite{chang2022maskgit}, MAGE~\cite{li2023mage}), the model predicts masked indices in parallel:
\begin{equation}
    \mathcal{L}_{\text{MAR}} = \mathbb{E}_{x \sim p(x),\, \mathcal{M} \sim \text{Mask}} \left[ \sum_{m \in \mathcal{M}} -\log p_\theta(s_m \mid s_{\mathcal{M}^c}) \right],
\end{equation}
where \( \mathcal{M} \subset \{1, \ldots, n\} \) is a randomly sampled mask and \( s_{\mathcal{M}^c} \) are the unmasked tokens. In inference we use Gumbel sampling and scale the logits by temperature. 

\section{Experiments}
\noindent\textit{\textbf{Dataset.}} 
The proposed method and baseline comparisons were implemented using cardiac MR short-axis (SAX) images from the UK BioBank dataset \cite{petersen2015uk}, featuring voxel spacing of $1.8 \times 1.8 mm$ (axial) and $8 mm$ (longitudinal).
The dataset comprised 11,360 training and each 1,420 evaluation and testing subjects. For training we randomly select a slice and timeframe as data augmenation and apply horizontal as well as vertical flips. All SAX images were preprocessed via center-cropping to $128 \times 128 $ pixel cardiac regions.

\noindent\textit{\textbf{Model Zoo Overview.}}  
We evaluate eight models: five diffusion-based (LDM-8~\cite{rombach2022high}, MDTv2-B2~\cite{gao2023mdtv2}, UViT-B2~\cite{bao2023all}, DiT-B2~\cite{peebles2023scalable}) and three autoregressive (Taming~\cite{rombach2022high}, MAGE~\cite{li2023mage}, MaskGIT~\cite{chang2022maskgit}). All models use their base variants with a patch size of 2 where applicable. Training details are provided in Appendix Table~\ref{tab:training-config-diffusion} and Table~\ref{tab:training-config-tokenvq}.

\noindent\textbf{\textit{Unconditional Image Generation.}} 
We train all generative models on an unconditional image generation task in their native representation spaces—continuous for diffusion models and discrete for autoregressive transformers. Each model generates 1,420 images, which we evaluate using generational Fréchet Inception Distance (FID) and Kernel Inception Distance (KID) based on Inception-v3 features~\cite{szegedy2016rethinking}. We also report generation time per sample (TPS) on an NVIDIA A6000 to assess computational efficiency.

\noindent\textbf{\textit{Reconstruction vs. Generation.}}
To evaluate the reconstruction–generation tradeoff, we apply each model to images with increasing mask ratios and assess performance using two complementary metrics: reconstruction Fréchet Inception Distance (rFID)~\cite{heusel2017gans} to measure perceptional quality relative to the ground truth, and Peak Signal-to-Noise Ratio (PSNR) for pixel-level accuracy. This approach systematically assesses how well each model balances local fidelity and global semantic plausibility as missing information grows.

\noindent\textbf{\textit{Temperature-aware Sampling.}}
We study controllability of the reconstruction–diversity tradeoff by varying sampling parameters across masking ratios in diffusion and autoregressive models. Plotting these effects reveals how adjusting sampling enables precise control over the balance between reconstruction accuracy and semantic diversity, especially as inpainting becomes more challenging with larger masked regions.

\begin{figure}[t]
\centering
\begin{minipage}[c]{0.55\textwidth}
    \centering
    \captionof{table}{Comparison of unconditioned image generation performance. Models are grouped by token types.}
    \resizebox{\linewidth}{!}{%
    \begin{tabular}{lllrr}
\hline
\textbf{Model} & \textbf{First Stage} & \textbf{Type} & \textbf{FID$\downarrow$} & \textbf{KID$\downarrow$} \\ 
\hline
\multicolumn{5}{l}{\textit{Pixel-based}} \\
ADM~\cite{dhariwal2021diffusion} & \multicolumn{1}{c}{/} & Diff & 29.15 & 0.0107 \\
\hline
\multicolumn{5}{l}{\textit{Vector-quantized tokens}} \\
Taming~\cite{esser2021taming} & VQGAN-F8 & AR & 58.69 & 0.0548 \\
MaskGIT~\cite{chang2022maskgit} & VQGAN-F8 & MAR & \underline{43.89} & \underline{0.0313} \\
MAGE~\cite{li2023mage} & VQGAN-F8 & MAR & 51.43 & 0.0441 \\
\hline
\multicolumn{5}{l}{\textit{Continuous-valued tokens}} \\
LDM-8~\cite{rombach2022high} & VAEKL-F8 & Diff & 40.13 & 0.0273 \\
MDTv2-B2~\cite{gao2023mdtv2} & VAEKL-F8 & Diff & 47.89 & 0.0368 \\
UViT-B2~\cite{bao2023all} & VAEKL-F8 & Diff & \textbf{38.78} & \textbf{0.0251} \\
DiT-B2~\cite{peebles2023scalable} & VAEKL-F8 & Diff & 39.53 & 0.0263 \\
\hline
\end{tabular}
    }
    \label{tab:image_gen_comparison}
\end{minipage}
\hfill
\begin{minipage}[c]{0.42\textwidth}
    \centering
    \includegraphics[width=\linewidth]{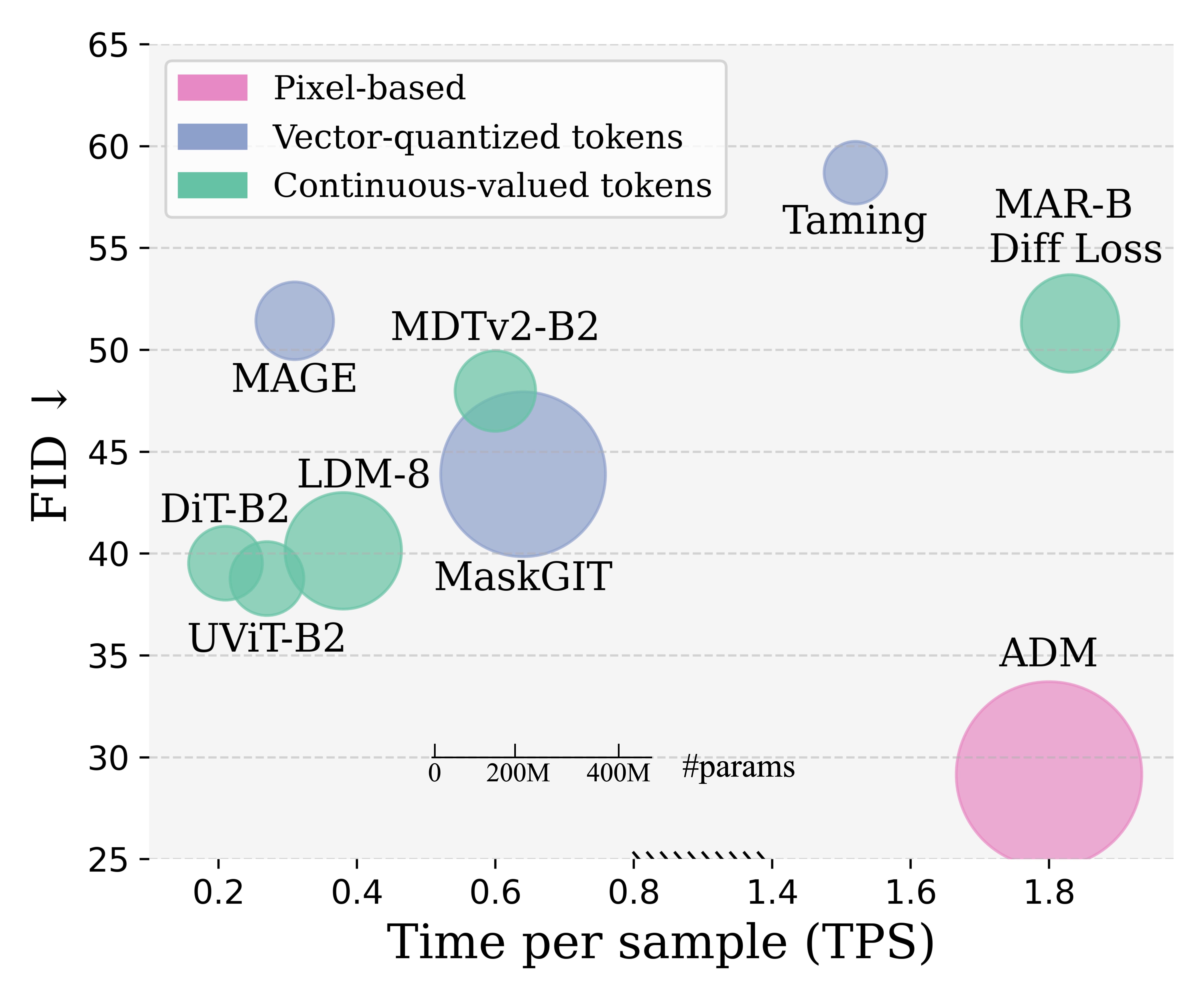}
    \caption{Comparison of models across three categories, plotting time per sample against FID. Dot sizes represent the number of trainable parameters.}
    \label{fig:bubble}
\end{minipage}
\end{figure}

\begin{figure}[ht]
    \centering
\includegraphics[width=0.75\textwidth]{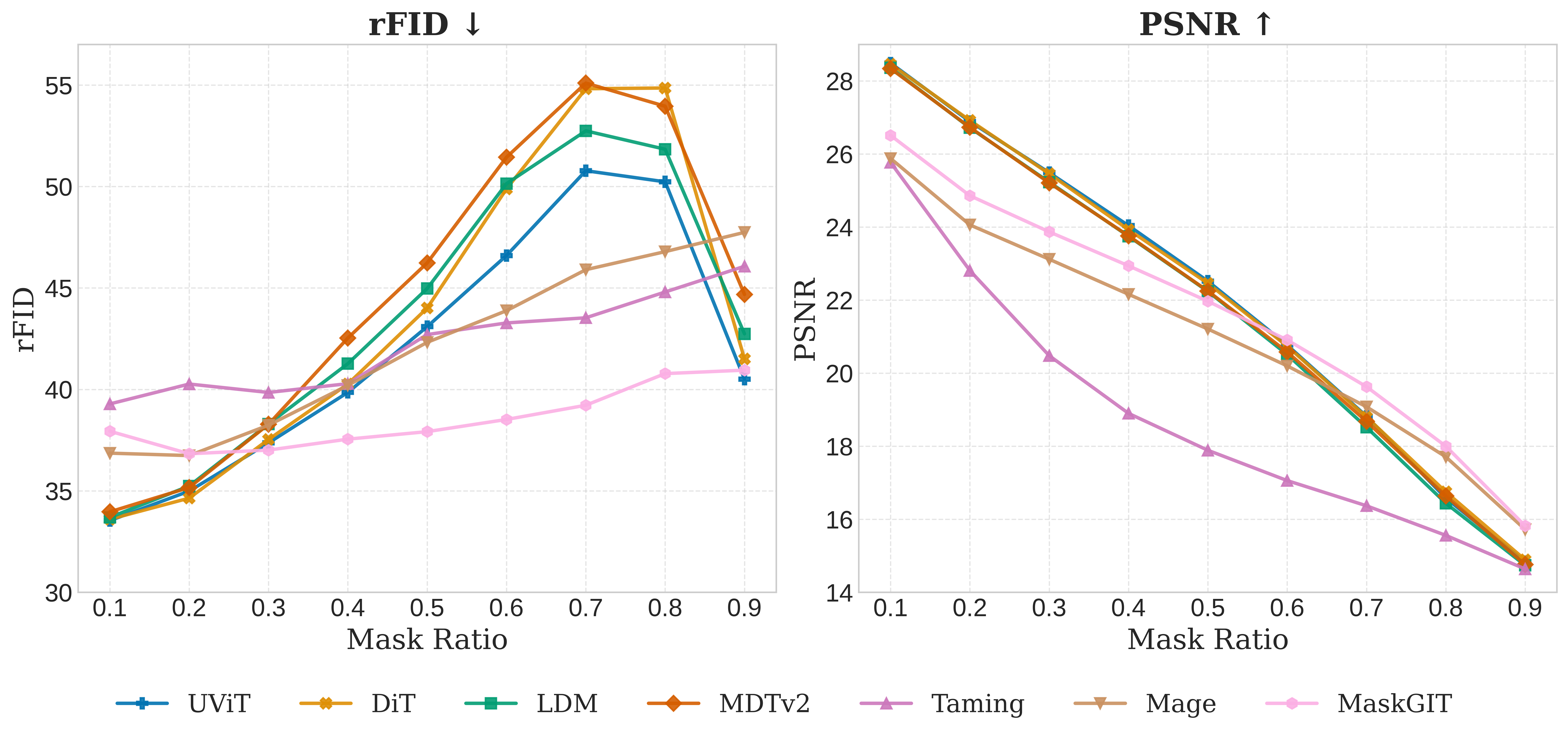}
    \caption{Development of rFID and PSNR metrics over increasing mask ratios for various generative models.}
    \label{fig:linear_generative}
\end{figure}

\begin{figure}[ht]
    \centering
\includegraphics[width=0.75\textwidth]{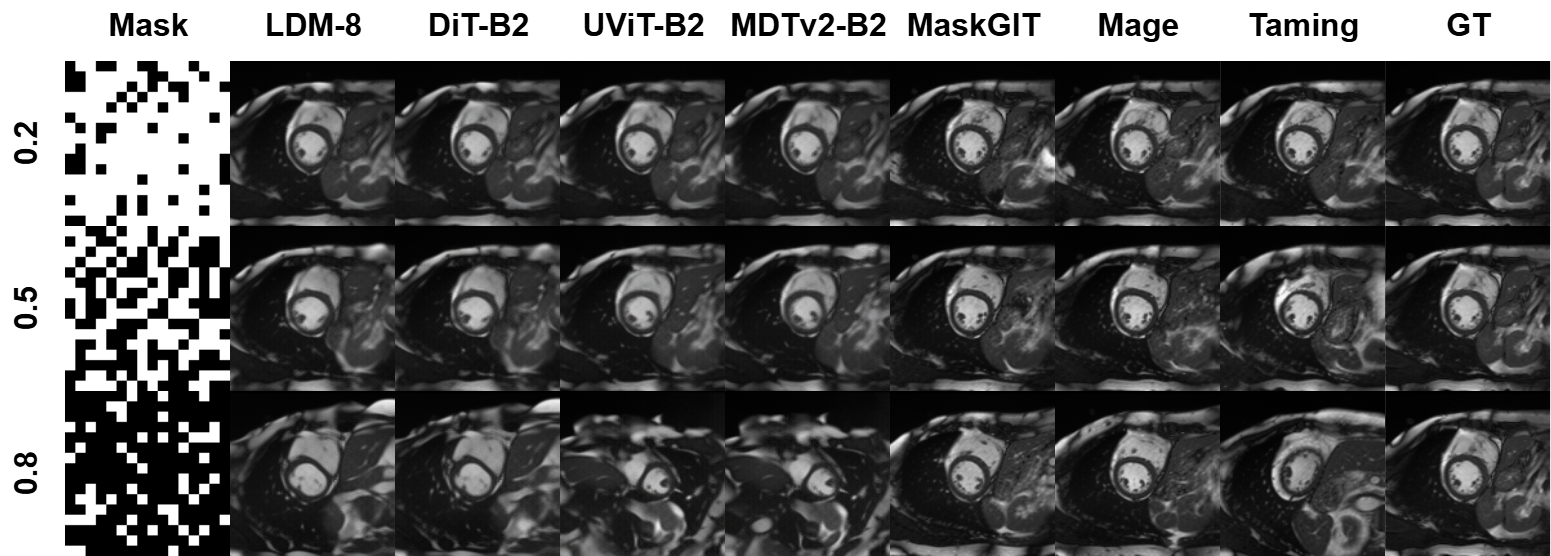}
    \caption{Qualitative comparison of generative models under varying masking ratios with a visual mask (black equals masked area) on the left and the ground truth (GT) on the right.}
    \label{fig:vis_generative}
\end{figure}



\section{Results and Discussion}

\noindent\textit{\textbf{Unconditional Image Generation.}}
Table~\ref{tab:image_gen_comparison} and Figure~\ref{fig:bubble} show that ADM, while offering strong quality, is slow and serves as a pixel-space proxy baseline. Diffusion models on continuous latents offer better quality–efficiency tradeoffs. Especially UViT-B2, with Vision Transformers and long skip connections, performs best, highlighting the benefits of global context and architectural shortcuts. DiT performs slightly worse, indicating diminishing returns from scaling in limited-data regimes. MDTv2-B2, using masked latent modeling, underperforms, suggesting contextual objectives are less effective. Among autoregressive models, MaskGIT achieves the best balance of speed and quality, thanks to its parallel, bidirectional prediction aligned with image structure. MAGE is faster but sacrifices fidelity. In contrast, causal, sequential models like Taming lack 2D priors, leading to slower, less coherent outputs.

\begin{figure}[t]
  \centering
  \resizebox{.75\textwidth}{!}{%
    \begin{tikzpicture}[font=\supersmall, scale=1]
    \node[anchor=north] at (0.7, 13.3) {(a)};
    \node[anchor=north] at (4.8, 13.3) {(b)};
    \node[anchor=south west, scale=1.2] at (0,10)
    {\includegraphics[width=3.3cm]{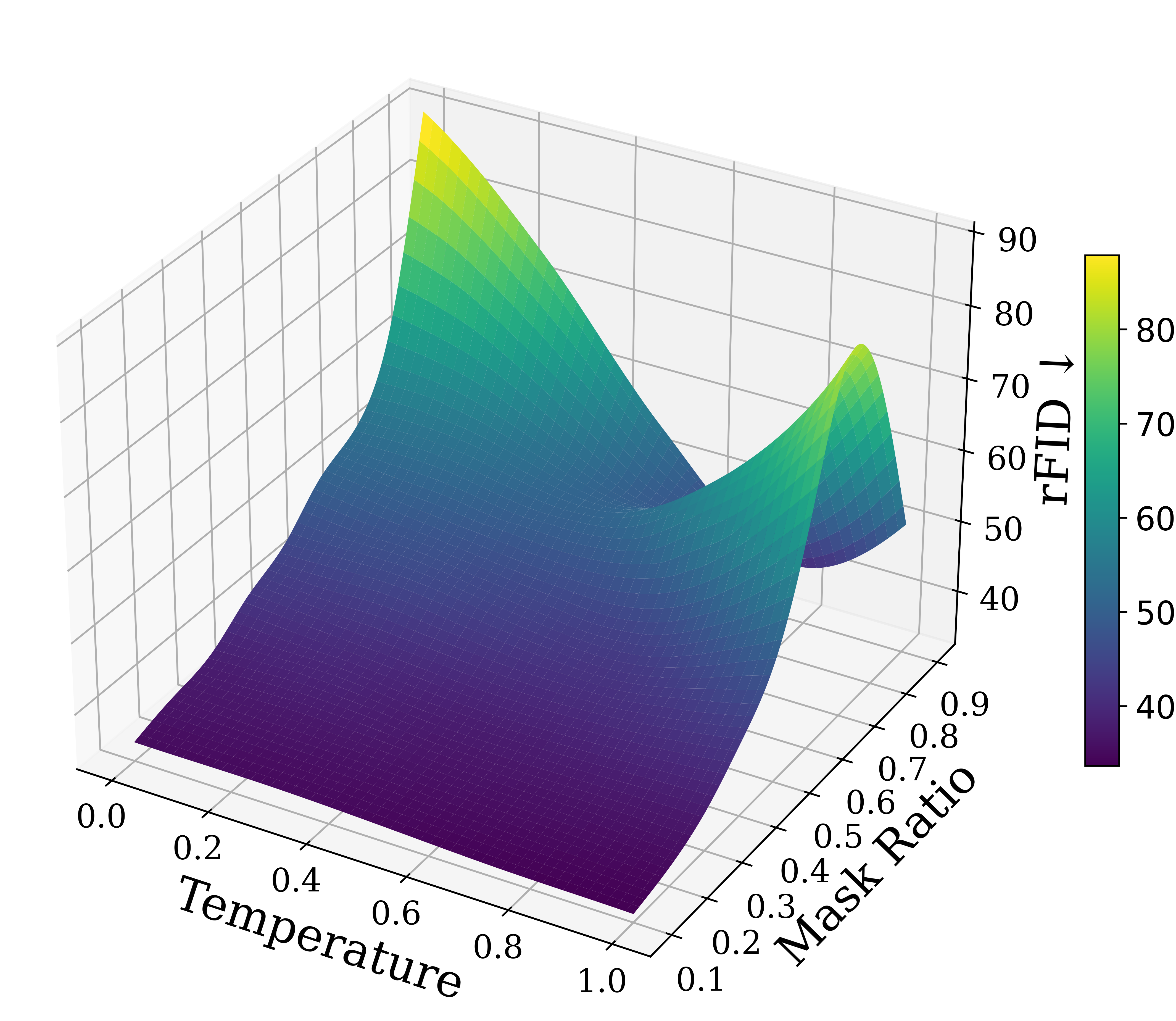}};
    \node[anchor=south west, scale=1.2] at (4,10)
    {\includegraphics[width=3.3cm]{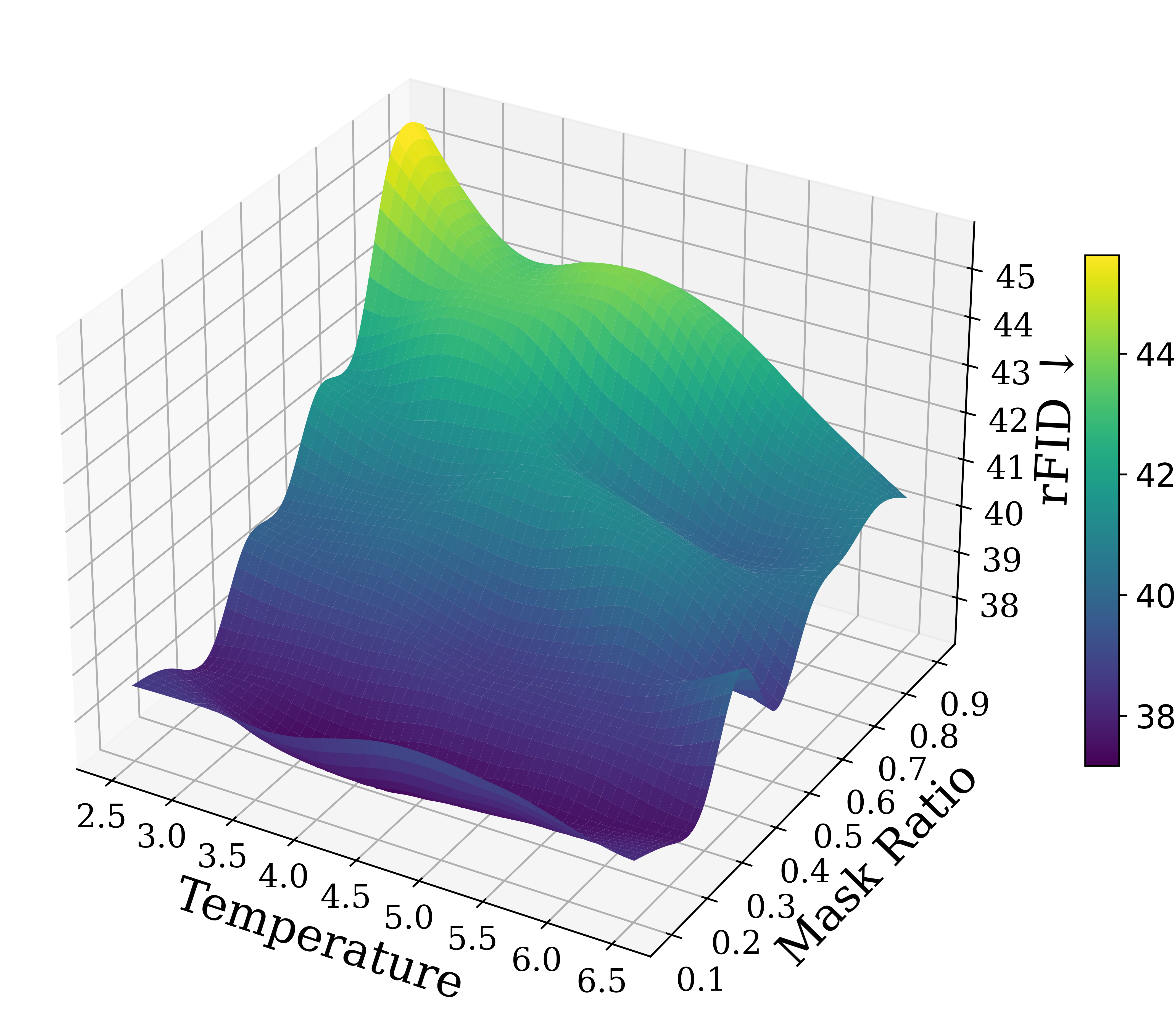}};

    \end{tikzpicture}
    }
  \caption{Visualizing the reconstruction-generation spectrum showing the rFID as function over mask ratio and temperature for UViT (a) and MaskGIT (b).}
  \label{fig:recon}
\end{figure}

\noindent\textit{\textbf{Diffusion Models Excel but Hallucinate with Limited Context.}}
Quantitative results in Figure~\ref{fig:linear_generative} show that diffusion models perform well in both perception and fidelity at low to moderate masking ratios, where sufficient context is available. However, as the masking ratio increases, perceptual quality declines—peaking in rFID at a ratio of 0.7 resulting in increased hallucination, as also illustrated in Figure~\ref{fig:vis_generative}. Interestingly, rFID improves again with higher masking rates, likely because the diffusion models switch to prior-driven generation, producing more coherent—though less faithful—outputs. As shown in Figure~\ref{fig:recon}, higher masking ratios require lower sampling temperatures to reduce hallucinations and preserve perception. We attribute the absence of strong priors for moderate masking to the generative training paradigm of diffusion models, which does not explicitly emphasize image reconstruction during training.

\noindent\textit{\textbf{Autoregressive Models Provide More Stable Perception but Lower Fidelity.}}
In contrast, autoregressive models follow a reconstruction-based training paradigm, explicitly learning to recover masked tokens through sequential, token-by-token generation. This process enforces strong local consistency by conditioning each token on previously generated ones, which leads to more stable rFID performance across masking ratios, as shown in Figures~\ref{fig:linear_generative} and~\ref{fig:vis_generative}. Their discrete token-based decoding and reconstruction losses prioritize accurate local recovery over perceptual diversity, enhancing robustness to context reduction. This is supported by Figure~\ref{fig:recon}, which shows that, unlike diffusion models, transformers require higher sampling temperatures to compensate for their lower inherent perceptual diversity. However, this strong focus on local reconstruction limits fine-grained fidelity when sufficient context is available, resulting in lower PSNR than diffusion models at lower masking ratios. At higher masking ratios, this bias toward precise token reconstruction helps maintain closer alignment with the original image, improving fidelity compared to diffusion models relying more heavily on learned priors.

\section{Conclusion}
Our study demonstrates that medical image reconstruction and generation are not separate tasks but lie on a continuous spectrum influenced by undersampling (masking ratio). We find that diffusion models excel in scenarios with sufficient context, producing high-fidelity and perceptually convincing outputs, but tend to hallucinate when context is limited, requiring careful temperature control. Conversely, autoregressive transformers, particularly MaskGIT-like models, maintain more stable perceptual quality across masking levels due to their explicit reconstruction training and token-based generation, though at the cost of reduced fidelity under greater masking ratios. By modularizing compression and generative stages into a flexible “generative models zoo”, we systematically evaluated these architectures and confirmed that MaskGIT-like transformers offer the most robust tradeoff between fidelity, perception, and computational efficiency across the full reconstruction–generation spectrum. These insights provide a nuanced understanding of how model design and training paradigms impact the balance between accurate reconstruction and realistic generation, informing the development of tailored solutions for diverse medical imaging tasks.

\clearpage  


\appendix
\newpage


\section{Appendix}
\subsection{Stage 1: Tokenizer Results}
\begin{table}[ht]
\centering
\begin{tabular}{lccccc}
\hline
\textbf{Model} & \textbf{rFID} ↓ & \textbf{rKID} ↓ & \textbf{PSNR} ↑ & \textbf{SSIM} ↑ & \textbf{LPIPS} ↓ \\
\hline
VQGAN-F8  & 38.74 & 0.0279 & 28.42 & 0.824 & 0.047 \\
VAEKL-F8  & 34.24 & 0.0243 & 30.29 & 0.887 & 0.062 \\
\hline
\end{tabular}
\caption{Performance comparison of different autoencoders used. Arrows indicate whether lower (↓) or higher (↑) values are better.}
\label{tab:autoencoder_comparison}
\end{table}
\subsection{Training Configurations}

\begin{table}[h!]
\centering
\caption{Training configuration settings by component (diffusion-based models)}
\begin{tabular}{lccccc}
\hline
\textbf{Component} & \textbf{MDTv2-B2} & \textbf{ADM} & \textbf{UViT-B2} & \textbf{DiT-B2} & \textbf{LDM} \\
\hline
\multicolumn{6}{c}{\textbf{Diffusion}} \\
\hline
Steps & 1000 & 1000 & 1000 & 1000 & 1000 \\
Predict $\sigma$ & true & true & false & true & false \\
Objective & noise & noise & noise & noise & noise \\
Schedule & linear & linear & linear & linear & linear \\
\hline
\multicolumn{6}{c}{\textbf{Optimizer}} \\
\hline
Name & Adan & AdamW & AdamW & AdamW & AdamW \\
LR & 0.0003 & 0.0001 & 0.0001 & 0.0001 & 0.0001 \\
Betas & [0.98, 0.92, 0.99] & [0.9, 0.999] & [0.9, 0.99] & [0.9, 0.95] & [0.9, 0.95] \\
Weight Decay & 0.0 & 0.01 & 0.03 & 0.01 & 0.01 \\
\hline
\multicolumn{6}{c}{\textbf{Scheduler}} \\
\hline
Type & / & / & Linear & Linear & / \\
Warmup Steps & / & / & 5000 & 5000 & / \\
\hline
\end{tabular}
\label{tab:training-config-diffusion}
\end{table}

\begin{table}[h!]
\centering
\caption{Training configuration settings by component (MaskGIT, Mage, Taming)}
\begin{tabular}{lccc}
\hline
\textbf{Component} & \textbf{MaskGIT} & \textbf{Mage} & \textbf{Taming} \\
\hline
\multicolumn{4}{c}{\textbf{Mask Generator}} \\
\hline
Distribution & $\mathcal{N}\left(0.5,0.25^2\right)$ & $\mathcal{N}\left(0.55,0.25^2\right)$ & / \\
Truncation & [0.05,0.95] & [0.5,1.0] & / \\
\hline
\multicolumn{4}{c}{\textbf{Optimizer}} \\
\hline
Name & AdamW & AdamW & Adam \\
LR & 0.0001 & 0.0001 & 0.0001 \\
Betas & [0.9, 0.96] & [0.9, 0.95] & [0.9, 0.999] \\
Weight Decay & 0.01 & 0.01 & 0.01 \\
\hline
\multicolumn{4}{c}{\textbf{Scheduler}} \\
\hline
Type & Linear & Linear & / \\
Warmup Steps & 5000 & 5000 & / \\
\hline
\end{tabular}
\label{tab:training-config-tokenvq}
\end{table}

\end{document}